# Photochemical reaction enabling the engineering of photonic spin-orbit coupling in organic-crystal optical microcavities

Qian Liang[1], Xuekai Ma[2], Jiahuan Ren[1,3], Teng Long[1], Chunling Gu[4], Cunbin An[1], Hongbing Fu[1,*], Stefan Schumacher[1,5,6], Qing Liao[1,*]

[1]Beijing Key Laboratory for Optical Materials and Photonic Devices, Department of Chemistry, Capital Normal University, Beijing 100048, China
[2]Department of Physics and Center for Optoelectronics and Photonics Paderborn (CeOPP), Paderborn University, 33098 Paderborn, Germany
[3]Hebei Key Laboratory of Optic-Electronic Information Materials, College of Physics Science and Technology, Hebei University, Baoding, 071002, PR China
[4]Institute of Process Engineering, Chinese Academy of Sciences, Beijing, 100190, China
[5]Institute for Photonic Quantum Systems (PhoQS), Paderborn University, 33098 Paderborn, Germany
[6]Wyant College of Optical Sciences, University of Arizona, Tucson, Arizona 85721, United States

E-mail: hbfu@cnu.edu.cn; liaoqing@cnu.edu.cn

**Abstract**

The control and active manipulation of spin-orbit coupling (SOC) in photonic systems is fundamental in the development of modern spin optics and topological photonic devices. Here, we demonstrate the control of an artificial Rashba-Dresselhaus (RD) SOC mediated by photochemical reactions in a microcavity filled with an organic single-crystal of photochromic phase-change character. Splitting of the circular polarization components of the optical modes induced by photonic RD SOC is observed experimentally in momentum space. By applying an ultraviolet light beam, we control the spatial molecular orientation through a photochemical reaction and with that we control the energies of the photonic modes. This way we realize a reversible conversion of spin-splitting of the optical modes with different energies, leading to an optically controlled switching between circularly and linearly polarized emission from our device. Our strategy of in situ and reversible engineering of SOC induced by a light field provides a promising approach to actively design and manipulate synthetic gauge fields towards future on-chip integration in photonics and topological photonic devices.

**Introduction**

In condensed matter systems, spin-orbit coupling (SOC) of electrons is responsible for a number of spintronic and topological phenomena(*1*). In solid-state systems with broken inversion symmetry, SOC brings about the so-called Brychkov-Rashba(*2*) and Dresselhaus(*3*) Hamiltonians that underly revolutionary applications in optoelectronics including information processing and quantum optics(*4, 5*). In analogy with electronic systems, similar SOC phenomena are also observed in photonic systems(*6-11*) including metamaterials, optical waveguides, and liquid crystals, where the polarization of photons can replace the role of electron spins. In planar microcavities, the combination of the natural splitting between transverse electric (TE) and transverse magnetic (TM) modes and breaking inversion symmetry caused by in-plane optical anisotropy leads to an artificial non-Abelian gauge field for photons(*12*) (). Topological phenomena based on the emerging optical analogs of SOC have been intensely investigated in recent years(*13-17*). These photonic systems have been demonstrated to be a promising platform to realize synthetic Hamiltonians for cavity photons in an effective magnetic field(*18-20*). However, the vector potential describing the gauge fields typically remains unchanged due to the immutability of the device structures. Recently, K. Rechcińska *et al.* engineered artificial SOC in a liquid-crystal-filled microcavity with an applied electric field and realized a persistent photonic spin texture (i.e., left- and right-handed circular polarization splitting)(*11*), a characteristic feature of Rashba-Dresselhaus (RD) SOC with equal Rashba and Dresselhaus coupling strength. L. Zhang *et al.* applied a magnetic field to control this effect in perovskite ferroelectrics, which allows in situ adjustment of the generated persistent spin spiral(*21*). However, the persistent application of electric and magnetic fields inevitably leads to large energy consumption and are not conducive to on-chip integration for advanced all-optical circuits on a micro- or even nanoscale.

Manipulation with light fields, as an alternative approach, provides a powerful and easily controllable toolbox for engineering SOC due to its advantages including ultrafast, non-contact spatial and temporal control(*22-24*). More importantly, optical manipulation can be spatially structured on a microscale, which benefits its applications in on-chip integration and performance regulation of microdevices. Photochemical reactions have long been investigated to regulate chemical and physical properties of organic molecules and highly-ordered molecular crystals(*25-28*). Especially for organic single-crystals, molecular rearrangement or dimerization induced by photochemical reaction results in a change of optical anisotropy. Recently, we have demonstrated that organic single-crystals with optical anisotropy can be used to construct synthetic RD Hamiltonians and realize persistent photonic spin textures due to their distinct anisotropy of the refractive-index distribution(*16, 29-32*). In combination with photochromism, organic single-crystals offer the potential to dynamically change their molecular arrangement when the photochemical reactions occur, thus influencing the anisotropic distribution of the refractive index. Therefore, by combining the advantages of photoresponsiveness and engineering of pseudospin textures, organic single-crystal-filled microcavities may open a new avenue for in situ optical engineering of artificial

Hamiltonians. So far, to our knowledge, the reversible optical control of photonic SOC in organic crystal-filled microcavities remains unexplored.

Here, we prepare organic microcavities filled with photochromic molecular single-crystals of 9-anthracenecarboxylic acid (9AC, Figure 1a). The circular polarization splitting of the optical modes induced by the photonic RD effect is observed experimentally in momentum space. By applying an external ultraviolet light beam, we control the spatial molecular dimerization and induce a change in the anisotropic refractive index profile. This way we can modify the energies of orthogonally linearly polarized modes and bring different neighboring modes into resonance. This strategy leads to optically controlled reversible engineering of SOC and provides a new approach to manipulate synthetic gauge fields and RD SOC towards future on-chip integrated in photonics and topological photonic devices.

**Photochemical reaction of the material and the principle of Rashba-Dresselhaus spin-orbit coupling**

Figure 1a shows a schematic diagram of our organic microcavity. The 9-anthracenecarboxylic acid (9AC) microribbon is sandwiched between two mirrors of the silver films with thicknesses of 100 (±5) nm (reflectivity ≥ 99%) and 35 (±2) nm (reflectivity about 50%), respectively. The scanning electron microscope (SEM) measurements show the microcavity structure in Figure 1b. The SEM image of the cross-section of this microcavity shows clear two metallic silver layers separated by the organic layer with typical thickness of 830 nm.

The 9AC crystal acts as an optically uniaxial medium with high birefringence characteristics due to the long-range highly-ordered molecular stacking arrangement in the ribbons. The initial orientation of the 9AC molecules is determined by the intrinsic structure of the crystals and the principle of minimum energy. According to the 9AC crystalline database (CCDC: No. 863458), 9AC molecules stack preferentially along the crystal [100] direction (that is Y-direction in Figure 1a) with intermolecular spacing of 3.897 Å (left panel of Figure 1c). The 9AC molecules in the crystals can undergo a reversible [4+4] photodimerization reaction when irradiated by ultraviolet (UV)-light(*28, 33*). Thus, two neighboring molecules in the densest packing direction are conducive to photodimerization which generates dimers through an excimer-state (right panel of Figure 1c). The spacing of two adjacent dimers is then changed from 3.890 Å for monomers to 7.742 Å according to the crystalline database for 9AC dimers (CCDC: No. 128359). Due to the steric repulsion caused by this atypical "head-to-head" packing, the photoinduced dimers automatically revert back to the monomer state without further (UV-light) illumination.

For the initial orientation of the 9AC molecules in the crystals, the optically long axis of the 9AC molecule is oriented along the X-direction, as shown in Figures 1a and 1d. In such an arrangement, the effective refractive index for light polarized linearly along the X-direction is equal to the $n_x$ refractive index. For the Y-direction, the effective refractive index is equal to n_y. The value of the refractive index components is strongly related to the density of molecular stacking in the respective directions and orientation of transition dipoles. The largest refractive index is along the Y-direction because of the densest molecular stacking and the transition-dipole-moment orientation. At normal incidence, for a cavity mode with *l* antinodes, the corresponding cavity modes in both orthogonal linear polarizations, denoted by X$_l$ and Y$_l$, are separated in energy due to the difference of effective refractive indices, which is schematically shown in three-dimensional (3D) reciprocal space tomography in Figure 1f (green for the Y-polarized mode and grey for the X-polarized mode).

Upon illumination with UV-light, the photochemical reaction of monomers results in formation of 9AC dimers and increases the intermolecular distances (Figure 1e). The effective refractive index for X-polarized light is only slightly changed while it significantly decreases for Y-polarized light. As a result, the birefringence characteristics of 9AC microcrystals significantly changes upon application of UV-light, which brings about a change of energies of two linearly polarized modes. When the X-polarized mode and Y-polarized one with different parity (such as X$_l$ and Y$_{l\pm1}$) are close to degeneracy, the RD SOC appears(*11*) and persistent a spin helix forms when Rashba and Dresselhaus terms are equal (Figure 1f, blue for the right-handed circularly-polarized mode and red for the left-handed circularly-polarized mode). Theoretically, the RD SOC in the basis of circular polarizations can be described by an effective 2×2 Hamiltonian:

$$H(\boldsymbol{k}) = \begin{pmatrix} E_0 + \frac{\hbar^2}{2m}\boldsymbol{k}^2 - 2\xi k_y & \beta_0 + \beta_1 \boldsymbol{k}^2 e^{2i\varphi} \\ \beta_0 + \beta_1 \boldsymbol{k}^2 e^{-2i\varphi} & E_0 + \frac{\hbar^2}{2m}\boldsymbol{k}^2 + 2\xi k_y \end{pmatrix} \quad (1)$$

Here, $E_0$ is the energy at $\boldsymbol{k}=0$ and $m$ is the effective mass of cavity photons, $\boldsymbol{k} = (k_x, k_y)$ is the in-plane wave vector and $\varphi$ ($\varphi \in [0, 2\pi]$) is the in-plane polar angle. $\beta_0$ denotes the energy splitting of orthogonally linearly polarized modes of opposite parity at $k=0$ (here, we define it as $\beta_0 = E_Y - E_X$, where $E_X$ and $E_Y$ are the ground state energies of X-polarized and Y-polarized modes of different parity). $\beta_1$ is the strength of the intrinsic TE-TM splitting of the cavity modes at $k \neq 0$, and $\xi$ is the RD coupling strength. If we consider the case of the initial 9AC microcrystals, when $\beta_0 = 0$ for the cavity modes of X$_l$ and Y$_{l+1}$, these two orthogonally linearly polarized modes are resonant at $k=0$ and split into circularly polarized modes (Figure 1f). With the applied UV-light irradiation, the change of the effective refractive indices of n$_x$ and n$_y$ induces different energy shifts of two cavity modes. As a consequence, the degeneracy of these two modes is lifted and the RD spin texture disappears. This optically induced change in anisotropy of the effective refractive index, enabled by a molecular photochemical reaction, allows the engineering of a spin-orbit synthetic Hamiltonian for an organic crystal-filled optical cavity.

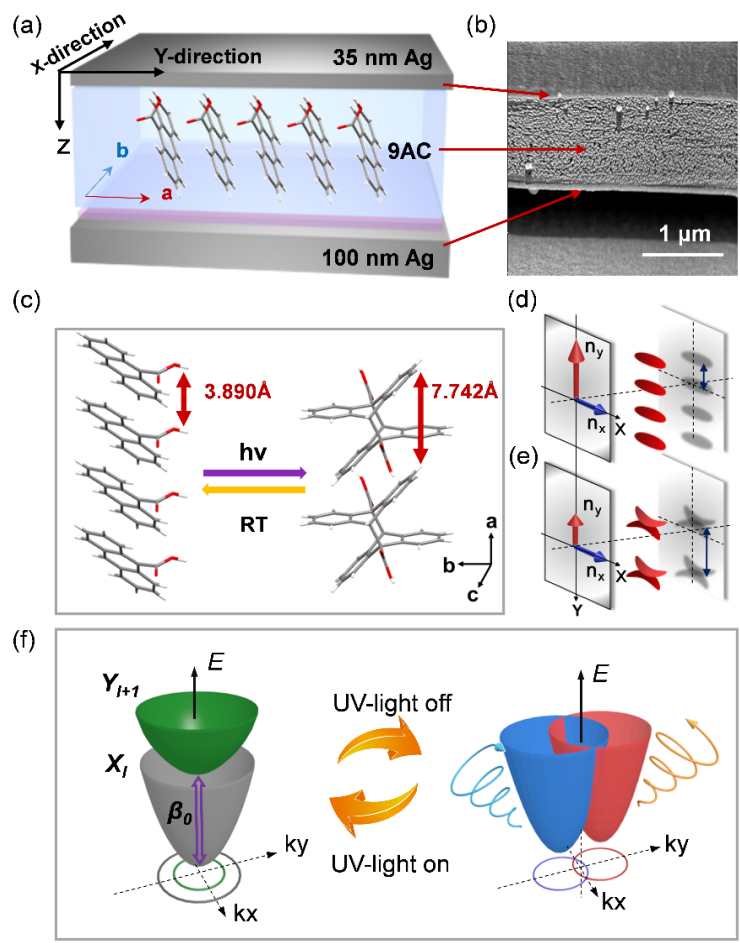

**Figure 1. The 9AC ribbon-filled optical microcavity, its photochemical reaction and RD SOC.** (a) The scheme of 9AC ribbon-filled microcavity and (b) its SEM image. (c) The scheme of photodimerization and dissociation reactions in the 9AC microribbons along the direction of the densest molecular stacking. (d) For the initial orientation of 9AC molecules in the microribbons, the effective refractive index for X(Y)-polarized light is equal to the $n_y$ ($n_x$) refractive index. (e) With UV-light applied to the microribbon, the adjacent 9AC molecules undergo a dimerization photochemical reaction, which decreases the effective refractive index $n_y$ for Y-polarized light. (f) Schematic of the reversible light-induced conversion of dispersion cones, illustrating the transition between states without (left) and with (right) RD SOC and correspondingly different polarization characteristics of the emission.

To experimentally realize the theoretical prediction discussed above, we fabricated optical microcavities filled with photochromic 9AC microcrystals. These were prepared through a facile floating drop method (see Supplementary Materials). The photoluminescence (PL) microscopy (Figure 2a) and scanning electron microscopy (SEM, Figure 2b) images show that the as-prepared 9AC samples have ribbon-like structure with smooth outer surfaces and sharp edges and bright green-color emission. These 9AC microribbons possess a typical length of hundreds to thousands of microns, a width of about 60-80 μm and a height of about 830 nm, estimated from atomic force microscopy (AFM) results (Figure 2c). The sharp spots in selected area electron diffraction (SAED) imaging clearly reveal that these 9AC microribbons are single crystalline (right panel of Figure 2b). According to the lattice parameters of the monoclinic phase of 9AC crystals, such as $a = 3.897(3)$ Å, $b = 9.355(2)$ Å, $c = 28.980(3)$ Å, $\alpha = 90°$, $\beta = 90.79°$, and $\gamma = 90°$, the sets of SAED spots marked with circle and triangle are attributed to (100) and (010) Bragg reflections with $d$-spacing values of 3.88 and 9.34 Å, respectively. Moreover, the X-ray crystal diffraction (XRD) curve is dominated by a series of peaks corresponding to the crystal plane (001) with $d = 14.19$ Å, such as (002), (004), (008) peaks. This also suggests that the 9AC microribbons belong to a monoclinic crystal structure and adopt a lamellar structure with the crystal (001) plane parallel to the substrate.

Figure 2e shows the PL spectra of the 9AC microribbon under irradiation with 405-nm laser. At the initial time ($t = 0$), the PL emission exhibits a green band with the dominant peak at 523 nm. With increasing irradiation time, this green emission weakens and a new blue emission band at near 477 nm appears. Eventually, this blue emission band becomes dominant, resulting in a conversion of green-emissive 9AC microribbons to blue-emissive ones. Without the irradiation of the 405-nm laser, the blue-emissive microribbons quickly (details given below) recover to green-emissive ones. We demonstrate that this reversible process can be repeated six cycles without any damage to the crystals or detorioration of performance (Figure 2f). More details studies of fatigue resistance should be the subject of future work. To further verify the occurrence of the photochemical reaction, Raman spectra for two samples before and after irradiation were taken. As shown in Figure 2g, the Raman peaks in the range of 1350-1600 cm$^{-1}$, corresponding to the typical bands of the aromatic anthracene backbone, disappear after irradiation. This suggests that the vibrations of aromatic anthracene are significantly weakened by the photodimerization of the anthracene ring(*34*).

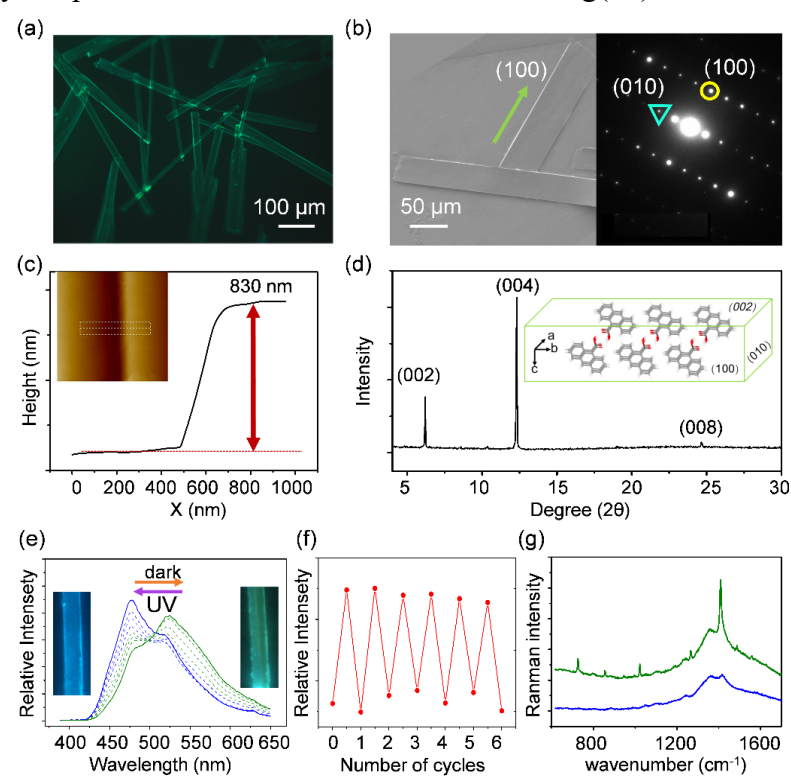

**Figure 2. The 9AC microcrystals and their photochemical reaction.** (a) The PL microscopy image of typical microribbons with the unfocused UV-light (330-380 nm). (b) SEM image and SAED pattern of typical microribbons. (c) The thickness and the corresponding AFM image of a typical microribbon. (d) XRD pattern for 9AC microribbons. Inset: side view of molecular packing within the microribbon. (e) Photochromic luminescence spectra as the increase (from green to blue lines) of the UV-light irradiation time and the corresponding fluorescence images of the microribbon. (f) Switch cycles (number of cycles) for the PL intensity of the blue emission upon the UV-light irradiation and subsequent recovery in the dark. (g) Raman spectra of the microribbon before (green) and after (blue) UV-light irradiation.

**Cavity modes and spin splitting**

We firstly perform the unpolarized angle-resolved reflectivity (ARR) of the microcavity with the 830-nm 9AC microribbon (hereinafter referred to as 9AC-MC) by using a home-made micro-scale ARR measurement setup (Figure S1) at room temperature. Figure 3a shows the two-dimensional (2D) reflectivity dispersion as a function of wavevector $k_y$ and two sets of modes with distinctive curvatures can be clearly seen. We have calculated these two modes by using the 2D cavity photon dispersion relations(*29*), which agrees well with experimental results (see the red and blue dashed lines in Figure 3a and Figure S2). The simulated refractive indices of two modes are 1.90 for X-polarized modes and 2.50 for Y-polarized ones, respectively, which supports the giant optical anisotropy of 9AC microribbons. In order to determine the polarization characteristics of cavity modes, the polarization-resolved ARR experiments of 9AC-MC have been carried out for the six different polarization components of light (horizontal-vertical, diagonal-anti-diagonal, and left and right circular)(*35*) to construct 3D tomography. Thus, the modes with larger and smaller curvatures of 9AC-MC are attributed to X- and Y-polarizations, corresponding to X- and Y-direction, respectively.

In the vicinity of 580 nm, a persistent spin helix induced by RD SOC emerges in this cavity and the dispersion curves are well fitted by the numerical results calculated by using Eq. (1) (the parameters used are given in the caption of Figure 4). We have extracted the Stokes vector components to analyze the polarization properties of the cavity modes. The $S_1$ components of the Stokes vector of $X_7$ and $Y_8$ branches present strongly linear polarization as shown in the dispersion relation of reflectivity, while their corresponding $S_3$ components are relatively weaker (Figure S3). In strong contrast, $X_6$ and $Y_7$ branches approach each other at $k_y = 0$ and resonate, which triggers spin splitting of the paraboloids along $k_y$ direction due to RD SOC. We also plotted the 2D wavevector map of the tomography (Figure 3b) and the corresponding Stokes vector components at 552 nm. As shown in Figures 3c and 3d, the Stokes component $S3$ is strongest and shows two separate circles with opposite signs, while the Stokes component $S1$ becomes very weak. We have also extracted Stokes vector components for the different wavelength (Figure S4). The experimental results indicate that $S3$ components become stronger and $S1$ components become weaker as the wavelength increases from 453 nm to 580 nm. This further demonstrates the occurrence of the RD SOC near 580 nm and the circular polarization splitting of the modes.

Figure 3e shows the ARR of the same 9AC-MC after irradiation with a 405-nm laser. The PL of this 9AC-MC has changed from green to blue (Figure S5), which indicates that the 9AC microribbon has undergone the photodimerization reaction. In agreement with expectations, the characteristic features of RD SOC in the dispersion plot have disappeared. Moreover, the position and curvature of one of the two sets of cavity modes have changed significantly. According to the theoretical simulation, the refractive indices for X-polarized and Y-polarized modes are determined to be 1.90 and 2.10, respectively. For the X-polarized mode, the effective refractive index almost remains constant while it has changed significantly due to the formation of dimers of 9AC molecules, which is consistent with our prediction. In this case, the 9AC molecules undergo a [4+4] photochemical reaction(*28, 33*) and form dimers, which is responsible for the modification of birefringence properties of 9AC-MC. We have also extracted the Stokes vector components to analyze the polarization property of the cavity modes. All these modes display strong linear polarization. Especially, the modes X6 and Y7 are separated and degenerate from resonant circular polarization to linear polarization, which are presented in 2D wavevector map of the tomography (Figure 3f-h).

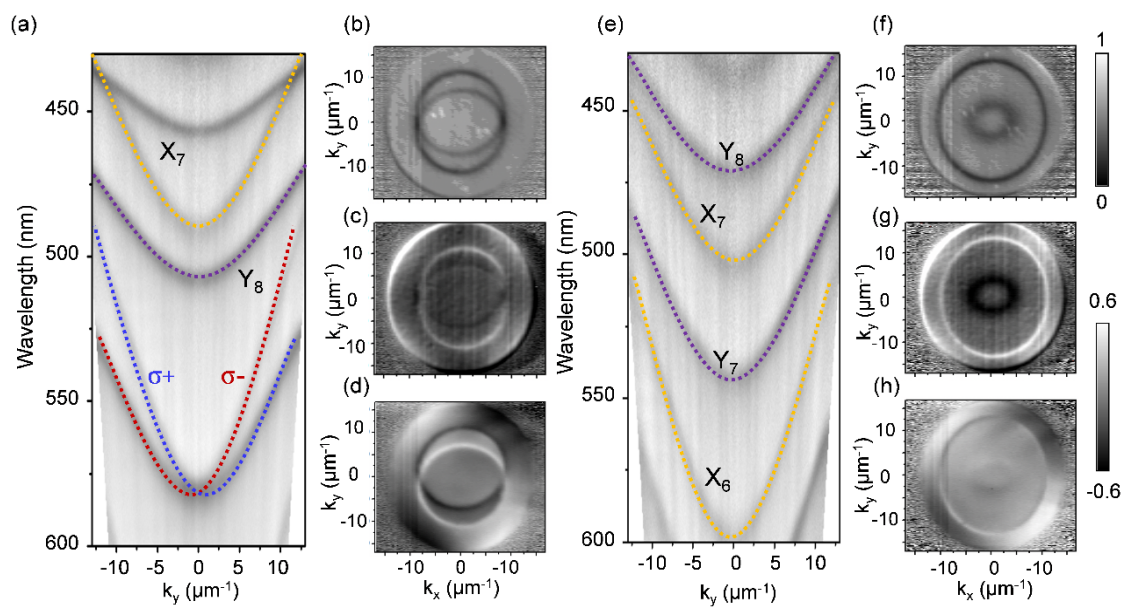

**Figure 3. Dispersion relation and cavity modes.** (a) ARR of the 9AC-MC with 830-nm 9AC microribbon. (b) The 2D wavevector map of the tomography and the corresponding Stokes vector components $S1$ (c) and $S3$ (d) at 552 nm. (e) ARR of the same 9AC-MC after the irradiation of 405-nm laser. (f) The 2D wavevector map of the tomography and the corresponding Stokes vector components $S1$ (g) and $S3$ (h) at 538 nm.

**Reversible engineering of Rashba-Dresselhaus spin-orbit coupling**

Taking advantage of the reversible photodimerization process of 9AC microribbons, in-situ and real-time engineering of RD SOC can be realized in our 9AC-MC. Figure 4a shows the ARR of the 9AC-MC after irradiation with the 405-nm laser for half an hour. As a result, there is no RD SOC because the orthogonally linearly-polarized modes are separated from each other. After 8 minutes at room temperature, the energy spacing between X7 and Y8 (or X6 and Y7) branches decreases significantly (Figure 4b) because of the decrease of the energy of Y8 branch, which is induced by the gradual dissociation of the 9AC dimers. After 16 minutes, the modes X7 and Y8 approach each other at $k_y = 0$ ($\beta_0 = 0$) and resonate (Figure 4c). As a result, a spin helix is triggered due to RD SOC. The left-handed and right-handed circularly polarized modes can be observed in ARR and shown in Figure S6. As recovery time reaches 25 minutes, the decrease of the energy of Y8 branch continues and the value of $\beta_0$ is less than 0 at $k_y = 0$. At this time, the resonance condition for X7 and Y8 branches is broken and both X7 and Y8 branches revert to linearly polarized modes (Figure 4d). When the time exceeds 35 min, the energy spacing between X6 and Y7 branches reaches zero (i.e., $\beta_0 = 0$) at $k_y = 0$. The RD spin splitting between X6 and Y7 modes is triggered (Figure 4e) and the microcavity system recovers the original state due to full dissociation of the 9AC dimers. We note that the relatively long-time scales for the manipulation of RD SOC in our present work are due to statistical effects; dimerization has to be triggered in many molecules of the crystal. Generally, faster optical switching will be achieved for higher pumping-laser intensity. Also, the reversion of the photochemical reaction is expected to be much faster on a molecular level. The use of different photochromic molecular crystals could be envisioned for future work, enabling optically induced back- and forth switching, both potentially with high speeds(*36*).

We have theoretically simulated the cavity modes and their refractive indices for these five cases and the obtained effective refractive indices are shown in Figure S7. During the recovery process, the effective refractive index $n_x$ remains almost unchanged, while the effective refractive index $n_y$ gradually changes with increasing recovery time, which is agreement with the picture of photodimerization and dissociation mechanism for organic photochromic molecules.

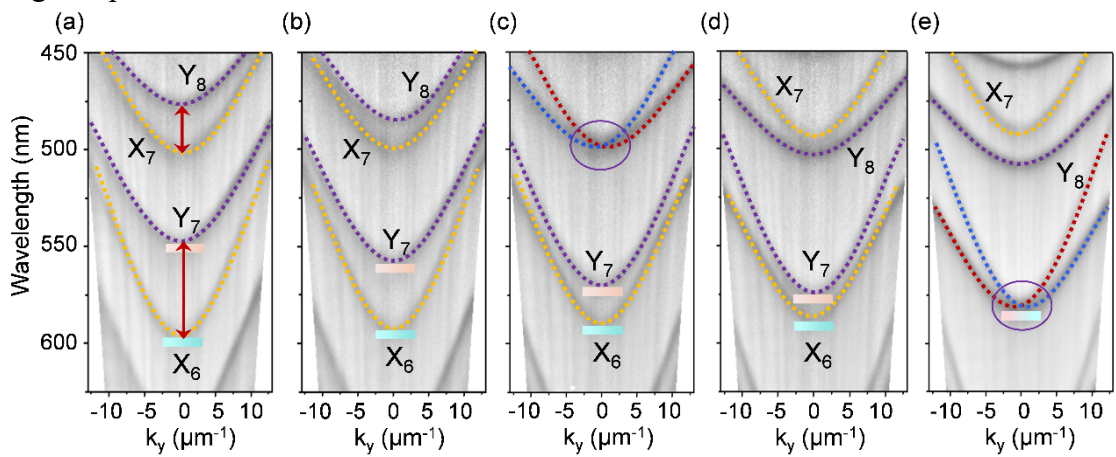

**Figure 4. In-situ reversible engineering of RD SOC.** (a-e) ARR of 9AC-MC after stopping irradiation for 0 min, 8 min, 16 min, 25 min and 35 min, respectively, and the corresponding simulated dispersion curves. The theoretical results of the spin splitting in (c,e) are calculated by using Eq. (1) with the parameters: $E_0 = 2.505$ eV (495 nm),

$m = 2.15 \times 10^{-5} m_e$ ($m_e$ is the free electron mass), $\beta_0 = 0$, $\beta_1 = 0.4$ meV μm$^2$, $\xi = 2.3$ meV μm for (c); $E_0 = 2.14$ eV (579 nm), $m = 1.85 \times 10^{-5} m_e$, $\beta_0 = 0$, $\beta_1 = 0.5$ meV μm$^2$, $\xi = 1.7$ meV μm for (e).

**Conclusion**

In summary, we demonstrate a light-controllable synthetic RD SOC in photochromic organic crystal-filled optical microcavities through a reversible photochemical reaction. The circular-polarization splitting of the optical modes induced by photonic RD SOC is observed experimentally in momentum space. By applying an external UV light, we control the molecular dimerization that induces a change of refractive index anisotropy. This way we achieve active modification of the energies of orthogonally linearly polarized modes and bring different modes into resonance. The resulting reversible engineering of SOC provides a new approach to manipulate synthetic gauge field and specifically RD SOC for the realization of future on-chip integrated photonics and topological photonic devices.

**Acknowledgements**

This work was supported by the National Natural Science Foundation of China (Grant No. 22150005, 22090022, and 22275125), the National Key R&D Program of China (2022YFA1204402, 2018YFA0704805 and 2018YFA0704802), the Natural Science Foundation of Beijing, China (KZ202110028043), the Science and Technology Innovation Program of Hunan Province (2022RC4039), Beijing Advanced Innovation Center for Imaging Theory and Technology. The authors thank Dr. HW Yin from ideaoptics Inc. for the support on the angle-resolved spectroscopy measurements.

# Supporting Information

**Photochemical reaction enabling the engineering of photonic spin-orbit coupling in organic-crystal optical microcavities**

Qian Liang[1], Xuekai Ma[2], Jiahuan Ren[1,3], Teng Long[1], Chunling Gu[4], Cunbin An[1], Hongbing Fu[1,*], Stefan Schumacher[1,5,6], Qing Liao[1,*]

[1]Beijing Key Laboratory for Optical Materials and Photonic Devices, Department of Chemistry, Capital Normal University, Beijing 100048, China
[2]Department of Physics and Center for Optoelectronics and Photonics Paderborn (CeOPP), Paderborn University, 33098 Paderborn, Germany
[3]Hebei Key Laboratory of Optic-Electronic Information Materials, College of Physics Science and Technology, Hebei University, Baoding, 071002, PR China
[4]Institute of Process Engineering, Chinese Academy of Sciences, Beijing, 100190, China
[5]Institute for Photonic Quantum Systems (PhoQS), Paderborn University, 33098 Paderborn, Germany
[6]Wyant College of Optical Sciences, University of Arizona, Tucson, Arizona 85721, United States

E-mail: hbfu@cnu.edu.cn; liaoqing@cnu.edu.cn

**Fabrication of single-crystal microribbons of 9-anthracenecarboxylic acid**

9-anthracenecarboxylic acid（9AC, TCI, > 97%) and ethyl acetate (Sigma, 99.5+%) were purchased from Innochem, respectively. They were directly used without further purification.

The 9AC microribbons were prepared by using the floating drop method. Firstly, 3.8 mg of 9AC was completely dissolved in 2.0 mL of filtered ethyl acetate. Next, ultrapure water was added into a Petri dish (diameter 60 mm), and then the obtained 9AC solution was slowly injected onto the surface of the purified $H_2O$. Finally, this Petri dish was covered and left in the dark for 48 hours. As the solvent of ethyl acetate evaporated, 9AC slowly crystallized and formed yellow ribbon-like microcrystals floating on the water surface.

**Microcavity fabrication**

Firstly, we used the metal vacuum deposition system (Amostrom Engineering 03493) to thermally evaporate silver film with the thickness of 85 (±5) nm (reflectivity: R ≥ 99%) on the glass substrate. The root mean square roughness ($R_q$) of this silver film in the 5 μm×5 μm area is 2.45 nm. A 20 (±2) nm $SiO_2$ layer was deposited using vacuum electron beam evaporate on the silver film with $R_q$ of 2.31 nm. The deposited rate was 0.2 Å/s and the base vacuum pressure was $3×10^{-6}$ Torr. Then the 9AC microcrystals were transferred to a silver/$SiO_2$ film substrate. Finally, 20 (±2) nm $SiO_2$ and 35 (±2) nm (R ≈ 50%) silver films were fabricated to form the microcavity. The $SiO_2$ layers are used to prevent the fluorescence quenching of the 9AC microbelt caused by directly contact of the metallic silver with the crystal.

**Optical Spectroscopy Method**

The reflectivity and photoluminescence spectroscopy were measured at room temperature by using a home-made micro-area Fourier image, which was presented to the spectrometer slit through four lenses. The schematic diagram of the angle-resolved experimental setup is shown in Scheme S1. The reflectivity spectrum was collected by the spectrometer with a 300 lines/mm grating and a 400×1340 pixel liquid nitrogen cooled charge-coupled (CCD). In order to investigate the polarization properties, we placed a quarter-wave plate, a half-wave plate and a linear polarizer in the detection optical path to resolve the circular polarization in the left-handed and right-handed (σ+ and σ-) basis.

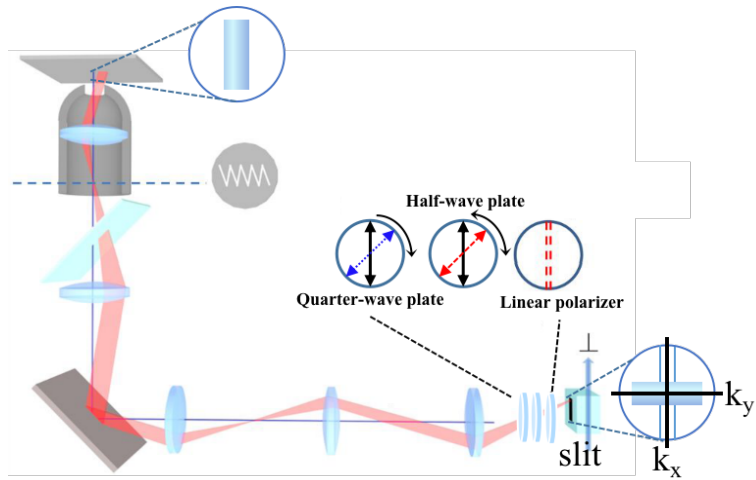

**Figure S1.** The schematic diagram of the angle-resolved measurement setup.

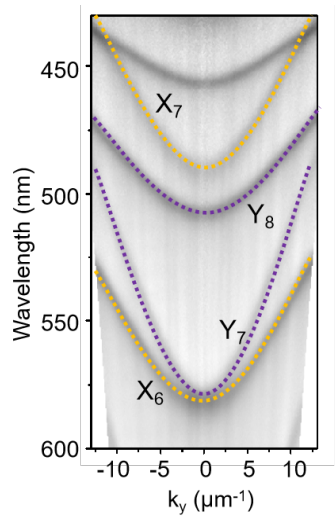

**Figure S2.** The simulated results for the Y-polarized and X-polarized cavity modes. The simulated refractive indices of the two modes are 2.50 and 1.90, respectively, corresponding to the red curve and blue curve.

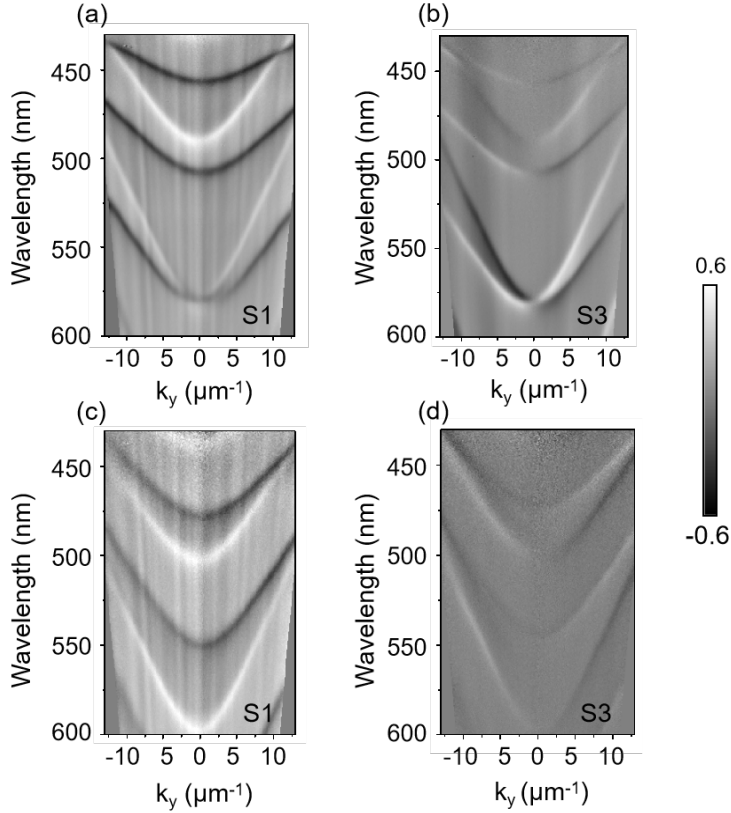

**Figure S3.** Stokes vector *S*1 (a) and *S*3 (b) of an optical microcavity filled with 830-nm 9AC microribbons. Stokes vector *S*1 (c) and *S*3 (d) of this microcavity after the irradiation of a 405-nm laser.

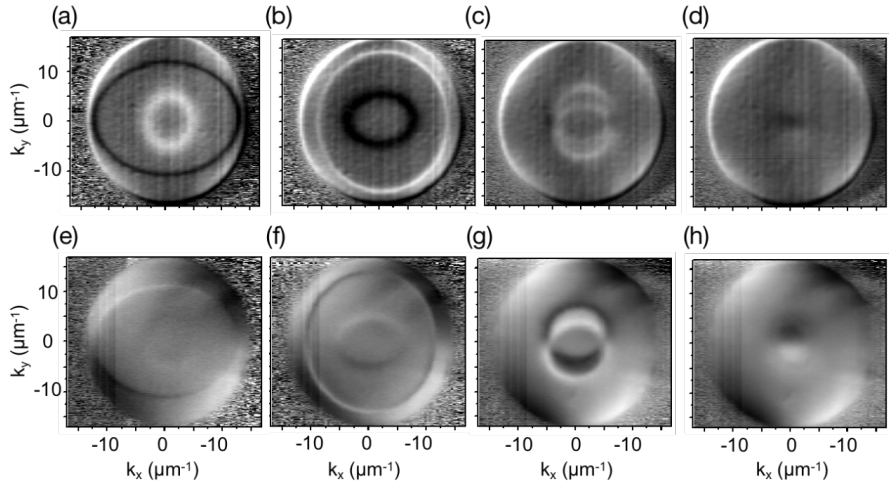

**Figure S4.** Cross-section plots of the 2D wavevector map of the tomography at different wavelengths in the momentum space. Stokes components *S*1 and *S*3 at 453 nm (a,e), 481 nm (b,f), 567 nm (c,g), and 580 nm (d,h), respectively.

As shown in Figure S4(a), two perpendicular and independent ellipses can be clearly seen in the Stokes *S*1 component at 453 nm, which correspond to the modes Y9 and X7, respectively. At the same time, their Stokes *S*3 component is very weak (Figure S4(e)), which indicates that Y9 and X7 modes are linearly polarized. Similarly, two perpendicular and independent ellipses (Figure S4(b,f)) at 481 nm, corresponding respectively to the modes Y8 and X7, also show strong linear polarization. In sharp contrast, the *S*3 component at 567 nm present two separate circles with opposite signs, that is, two opposite circular polarizations (Figure S4(g)), while the signals in the *S*1 component become very weak (Figure S4(c)), which indicates the occurrence of the RD SOC. At 580 nm, the signals in the Stokes component *S*3 are completely separated (Figure S4(h)) and in *S*1 cannot be distinguished (Figure S4(d)).

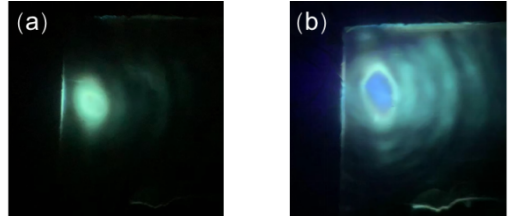

**Figure S5.** The PL microscopy image of the optical microcavity filled with 9AC microribbons before (a) and after (b) the irradiation of a 405-nm laser.

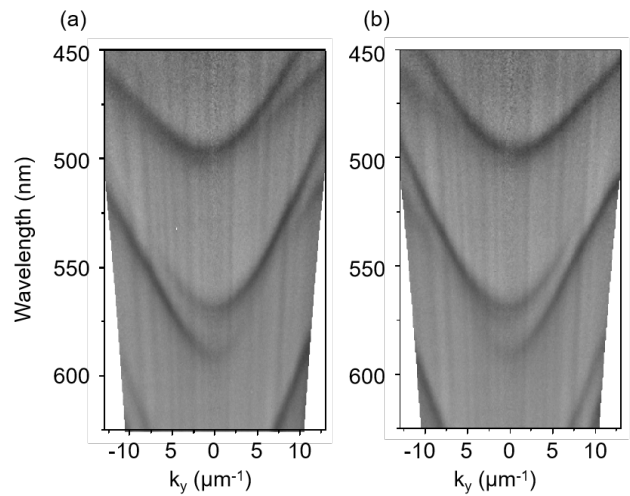

**Figure S6.** Angle-resolved left-handed (a) and right-handed (b) circularly polarized reflectivity of the optical microcavity filled with the 830-nm 9AC microribbon after stopping radiation for 16 min.

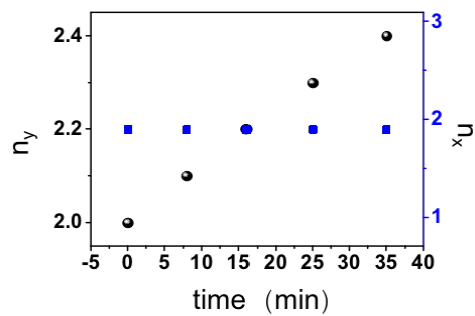

**Figure S7.** The simulated effective refractive indices of the 9AC-MC at different recover time.